\documentclass[preprint,12pt]{elsarticle}

\usepackage{graphicx}
\usepackage{amssymb}
\usepackage{dcolumn}
\usepackage{mathtools}
\usepackage{amsmath}
\usepackage{graphics}
\usepackage{natbib}
\usepackage[usenames, dvipsnames]{color}
\biboptions{square}
\journal{Solid State Communications}

\begin{document}

\begin{frontmatter}

\title{Quantum oscillations of the topological surface states in low carrier concentration crystals of  Bi$_{2-x}$Sb$_{x}$Te$_{3-y}$Se$_{y}$}

\author[uva]{Y. Pan\corref{cor1}}
\ead{y.pan@uva.nl}
\author[uva]{A. M. Nikitin}
\author[uva]{D. Wu\fnref{fnref1}}
\fntext[fnref1]{Present address: School of Physics and Engineering, Sun Yat-Sen University, Guangzhou, China}
\author[uva]{Y. K. Huang}
\author[ru]{A. Puri}
\author[ru]{S. Wiedmann}
\author[ru]{U. Zeitler}
\author[uva]{E. Frantzeskakis}
\author[uva]{E. van Heumen}
\author[uva]{M. S. Golden}
\author[uva]{A. de Visser}
\ead{a.devisser@uva.nl}
\cortext[cor1]{Corresponding author}

\address[uva]{Van der Waals - Zeeman Institute, University of Amsterdam, Science Park 904, 1098 XH Amsterdam, The Netherlands}
\address[ru]{High Field Magnet Laboratory, Radboud University, Toernooiveld 7, 6525 ED Nijmegen, The Netherlands}

\date{\today}

\begin{abstract}
We report a high-field magnetotransport study on selected low-carrier crystals of the topological insulator Bi$_{2-x}$Sb${_x}$Te$_{3-y}$Se$_{y}$. Monochromatic Shubnikov - de Haas (SdH) oscillations are observed at 4.2~K and their two-dimensional nature is confirmed by tilting the magnetic field with respect to the sample surface. With help of Lifshitz-Kosevich theory, important transport parameters of the surface states are obtained, including the carrier density, cyclotron mass and mobility. For $(x,y)=(0.50,1.3)$ the Landau level plot is analyzed in terms of a model based on a  topological surface state in the presence of a non-ideal linear dispersion relation and a Zeeman term with $g_s = 70$ or $-54$. Input parameters were taken from the electronic dispersion relation measured directly by angle resolved photoemission spectroscopy on crystals from the same batch. The Hall resistivity of the same crystal (thickness of 40~$\mu$m) is analyzed in a two-band model, from which we conclude that the ratio of the surface conductance to the total conductance amounts to 32~\%.
\end{abstract}

\begin{keyword}
A. Topological insulator
D. Quantum oscillations
D. Berry phase
E. Magnetotransport

\end{keyword}


\end{frontmatter}


\section{Introduction}

Topological insulators (TIs) in three dimensions (3D) attract much attention as versatile platforms to study new forms of quantum matter~\cite{Hasan&Moore2011}. TIs are bulk insulators with a non-trivial topology of the electronic bands that gives rise to metallic states at the surface~\cite{Hasan&Kane2010,Qi&Zhang2010}. The gapless surface states host a wealth of new physics, because they have a Dirac-type energy dispersion and the spin is locked to the momentum. As a result, they are immune to backscattering due to disorder, provided that time reversal symmetry is preserved. This makes TIs promising materials for applications in fields like spintronics and magnetoelectronics~\cite{Hasan&Kane2010,Qi&Zhang2010}. At the same time, TIs offer an almost unlimited source of test-case materials for new theoretical ideas and concepts, like the quantum spin Hall effect~\cite{Maciejko2011}, Majorana physics~\cite{Beenakker2013} and quantum computation~\cite{Nayak2008}.

Probably the best studied TI family consists of the layered compounds Bi$_2$Te$_3$, Bi$_2$Se$_3$, Sb$_2$Te$_3$, etc. The  prediction that these materials are 3D TIs with a single Dirac cone on the surface~\cite{Zhang2009}, was promptly verified in experiments by the surface sensitive technique of angle-resolved photoemission spectroscopy (ARPES)~\cite{Xia2009,Chen2009,Hsieh2009b}. However, the interior (bulk) of these workhorse TI materials is in general not a genuine insulator, because of the presence of charge carriers induced by impurities and defect chemistry. This seriously hampers the study of topological surface states in transport experiments, as well as potential device applications based on spin and charge transport. In order to solve this problem several research directions have been pursued, among which charge carrier doping ~\cite{Analytis2010a,Qu2010}, thin film engineering and electrostatic gating~\cite{Chen2010,Checkelsky2011}. Yet another route was promoted by Ren \textit{et al}.~\cite{Ren2011}, namely to approach the intrinsic topological insulator regime by optimizing the Bi$_{2-x}$Sb${_x}$Te$_{3-y}$Se$_{y}$ (in short BSTS) composition. The composition around ($x,y$)=(0.50,1.3) was found to be the optimum for bulk insulating behavior, as evidenced by a resistivity of several $\Omega$cm at liquid helium temperatures and a bulk carrier concentration of $\sim 2 \times 10^{16}$~cm$^{-3}$. The appealing topological properties of BSTS, notably a tunable Dirac cone, were furthermore demonstrated by ARPES~\cite{Arakane2012}, STM and STS~\cite{Ko2013} and THz Time Domain Spectroscopy~\cite{Tang2013}. We remark that recently also stoichiometric BiSbTeSe$_2$ has become an attractive material to investigate topological surface states~\cite{Segawa2012,Xu2014,Sulaev2015}.

Recently we reported an extensive magnetotransport study that aimed at the further investigation of the bulk-insulating properties of BSTS~\cite{Pan2014}. Single crystals with composition Bi$_{1.46}$Sb$_{0.54}$Te$_{1.7}$Se$_{1.3}$ produced the highest resistivity (12.6 $\Omega$cm) and lowest bulk carrier density ($0.2~\times 10^{16}$~cm$^{-3}$) at low temperatures. The contribution from the bulk and surface channels to the total resistance can be disentangled by employing a parallel resistor model. For a sample with a typical thickness of 100~$\mu$m, the ratio of the surface conductance over the total conductance is about 27~\%. Upon further reducing the sample thickness, this ratio increases and it can be as large as 97~\% for a $1~\mu$m thick sample~\cite{Pan2014}. The magnetoconductance of BSTS nanoflakes, prepared around the optimum composition, showed 2D weak antilocalization with an amplitude $\alpha \simeq -1$, as expected for transport dominated by topological surface states~\cite{Pan2014}.

In this paper we report a high-magnetic field transport study on selected, optimized BSTS crystals, which enabled us to probe the surface states by quantum oscillations in the resistance, via the Shubnikov - de Haas (SdH) effect. The SdH effect is a powerful tool to discriminate between 2D and 3D Fermi surfaces~\cite{Shoenberg1984}. At the same time, it may give direct access to the topological nature of the surface states via the geometric phase (Berry phase)~\cite{Taskin&Ando2011,Wright&McKenzie2013} of the quantum oscillations. Therefore, the SdH effect in TIs has received ample attention in the literature, notably through experiments carried out on bulk crystals of Bi$_2$Te$_3$, Bi$_2$Se$_3$ and Bi$_2$Te$_2$Se~\cite{Qu2010,Analytis2010a,Ren2010,Xiong2012,Petrushevsky2012}. In addition, the SdH effect was reported for non-stochiometric BSTS ($x,y$)=(0.50,1.3) bulk crystals~\cite{Taskin2011} and nanoflakes~\cite{Hsiung2013}. In most cases, the phase offset of the quantum oscillations obtained from a linear Landau level plot has been interpreted as a finite Berry phase. However, such an interpretation is not straight-forward because of the non-ideal Dirac dispersion and the sizeable Zeeman effect due to the large $g_s$-factor~\cite{Taskin&Ando2011,Wright&McKenzie2013}. Therefore, care should be taken when using the phase offset of the SdH oscillations as direct evidence for topological surface states.

Here we present SdH data for BSTS crystals with compositions Bi$_{1.5}$Sb$_{0.5}$Te$_{1.7}$Se$_{1.3}$ and Bi$_{1.46}$Sb$_{0.54}$Te$_{1.7}$Se$_{1.3}$. The SdH oscillations are monochromatic and their field-angular variation demonstrates their 2D nature. The standard analysis using Lifshitz-Kosevich theory gives a 2D carrier density of 1.5 and $0.8 \times 10^{12}$~cm$^{-2}$, and a cyclotron mass of 0.18 and 0.10$m_e$, respectively. For ($x,y$)=(0.50,1.3) the Landau level plot is analyzed in terms of a topological surface state with help of the non-linear dispersion relation measured directly by ARPES, and a Zeeman term with $g_s = 70$ or $-54$. In addition, we show that the Hall resistivity of the same crystal (of thickness 40~$\mu$m) can be analyzed successfully in a two-band model, from which a surface contribution to the total conductance of 32~\% is arrived at.

\section{Experimental}

The BSTS single crystals used for the SdH measurements were taken from the same batch that was prepared for the magnetotransport study reported in Ref.~\cite{Pan2014}. Here we focus on crystals around the optimum composition, $i.e.$ with $(x,y)=(0.50,1.3)$ and $(x,y)=(0.54,1.3)$, whereby the $x$- and $y$-values refer to nominal concentrations. For details of the single crystal growth procedure and characterization of the crystals by magnetotransport we refer to Ref.~\cite{Pan2014}. Flat rectangular samples were cut from the single-crystalline boule using a scalpel blade. Next the samples were cleaved in the $ab$-plane of the rhombohedral structure, at both top and bottom sides, using Scotch tape so as to obtain a thickness of around $50~\mu$m. The longitudinal, $R_{xx}$, and Hall resistance, $R_{xy}$, were measured in a  six-probe configuration. Current and voltage contacts were made by attaching thin (40~$\mu$m) copper wires to the crystals with silver paste. The exposure time to air between cleaving and mounting the samples in the cryostat was kept to a minimum of about one hour.

High magnetic fields were produced with a Bitter magnet ($B_{max} = 33$~T) at the High Field Magnet Laboratory at the Radboud University in Nijmegen. The samples were mounted on the platform of a mechanical rotator that could be cooled down to 1.7~K. The resistance was measured using a low frequency ac-technique with a SR830 DSP lock-in amplifier. The excitation current, $I$, flows in an arbitrary direction in the $ab$-plane and was typically 10~$\mu$A. Measurements were conducted for two polarities of the magnetic field, after which the longitudinal and Hall resistance were extracted by symmetrization. The field-sweep rate amounted to 30~mT/s. By rotating the sample platform, the angular variation of the SdH oscillations was determined. The magnetoresistance was always measured in the transverse configuration ($B \perp I$), since the rotation axis coincides with the direction of $I$.

ARPES measurements were performed at the SIS-HRPES endstation of the Swiss Light Source using a ScientaR4000 hemispherical electron analyzer. The data presented in this work were acquired at 16~K using 27~eV photons with linear horizontal polarization. The samples were cleaved and measured at a pressure better than $5 \times 10^{-11}$~mbar and the Fermi level position was determined using in-situ evaporated Au thin films that were in direct contact with the sample holder.

\section{Results and Analysis}

\subsection{Shubnikov - de Haas effect}

\begin{figure}
\includegraphics[height=15cm]{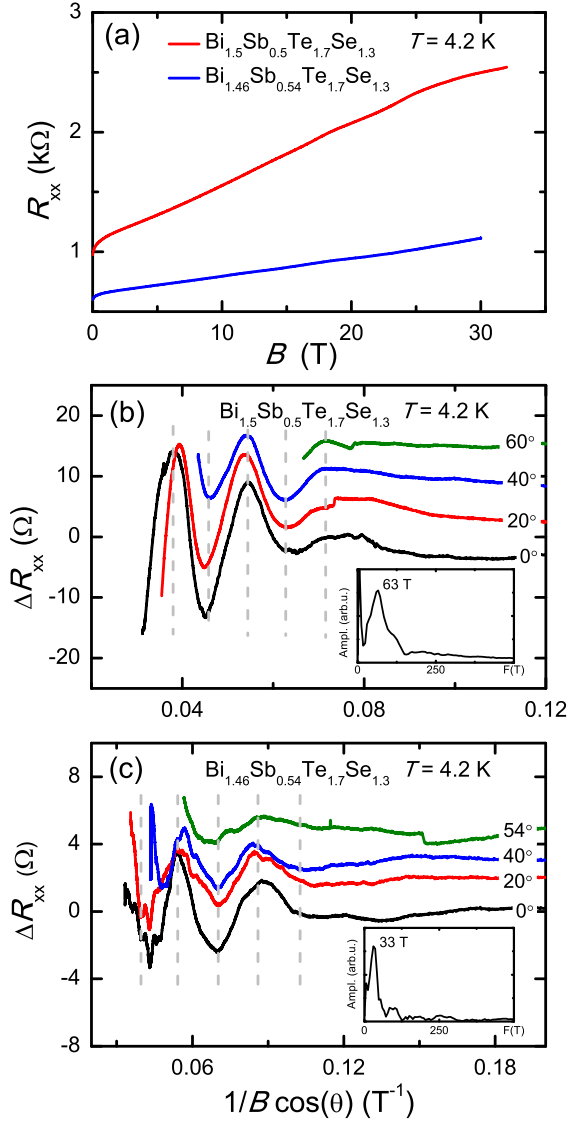}
\caption{Panel (a): Longitudinal resistance $R_{xx}$ of BSTS crystals with $(x,y) = (0.50,1.3)$ and $(x,y) = (0.54,1.3)$ as a function of the magnetic field at $T=4.2$~K and $\theta = 0^{\circ}$. Panel (b) and (c): Oscillatory component of the longitudinal resistance $\Delta R_{xx}$ plotted \textit{versus} $1/(B\cos\theta)$ for BSTS crystals with $(x,y) = (0.50,1.3)$ and $(x,y) = (0.54,1.3)$, respectively, at $T=4.2$~K. Here $\theta$ is the angle between the field and the normal to the sample surface ($c$-axis) and $B \cos \theta$ is the perpendicular component of the applied magnetic field. Curves for $\theta \neq 0$ are offset for clarity.  The positions of the minima and maxima of the SdH oscillations, marked by the vertical dashed lines, depend solely on $1/B \cos \theta$, which points to the 2D nature of the Fermi surface. The insets  show the fast Fourier transform of the data at $\theta = 0^{\circ}$.}
\end{figure}

We have measured the magnetoresistance of 10 different BSTS crystals in magnetic fields up to 30~T, applied along the rhombohedral axis ($c$-axis) at 4.2~K. In Fig.~1(a) we show the longitudinal resistance $R_{xx}$ of BSTS crystals with $(x,y) = (0.50,1.3)$ and $(x,y) = (0.54,1.3)$ as a function of the magnetic field at $T=4.2$~K and $\theta = 0^{\circ}$. After the initial sharp rise connected to the suppression of the weak antilocalization (WAL) in low fields~\cite{Pan2014}, the magnetoresistance increases in a quasi-linear manner without saturation. Selected crystals showed a clear SdH effect. These were further investigated to determine the temperature and angular variation of the SdH oscillations. While the magnetoresistance, $MR(B) = (R(B)-R(0))/R(0)$, typically has a magnitude of 100~\% near 30~T, the amplitude of the SdH signal is small and amounts to only 1~\% of the total resistance. After subtracting the smooth monotonic background contribution from $R_{xx}$ we obtain the SdH signal shown in Fig.~1(b) and 1(c). Here we trace $\Delta R_{xx}$ \textit{versus} $1/B$, in order to reveal the characteristic quantum oscillation period, where $\Delta R_{xx}$ refers to the difference between the oscillatory resistance in field and the smooth background. The data with the field perpendicular to the sample surface ($B \parallel c$-axis, $\theta = 0^{\circ}$) are given by the solid black lines. The angular variation of the SdH effect measured for angles $\theta \leq 60^{\circ}$ provides strong evidence the oscillations can be attributed to a 2D Fermi surface, since the positions of the minima and maxima at different $\theta$ coincide in the plots of $\Delta R_{xx}$ versus $1/(B \cos \theta$) as indicated by the vertical grey dashed lines in Fig.~1(b),(c). Here $\theta$ is defined as the angle between the field direction and the crystallographic $c$-axis. We remark that, strictly speaking, a 3D spheroidal Fermi surface (\textit{i.e.} an ellipsoid of revolution) with an aspect ratio such that the longer axis is along the reciprocal lattice vector $2 \pi /c$ could also be in agreement with the angular variation of the SdH data. However, this possibility can safely be excluded since the calculated 3D carrier density, $n_{3D}$, is at variance with the measured Hall data (see below). The fast Fourier transforms of the data at $\theta = 0^{\circ}$ give the SdH frequencies $F$ of 63$\pm 3$~T and 33$\pm 3$~T for $(x,y) = (0.50,1.3)$, and $(x,y) = (0.54,1.3)$, respectively. See the insets in Fig.~1(b) and 1(c). According to the Onsager relation, the extremal cross section of the Fermi surface, $A_{k} (E_{F})$, is proportional to the frequency, $F$, via the relation $A_{k} (E_{F} )= (2 \pi e / \hbar) \times F$, where $\hbar$ and $e$ are Planck's constant divided by $2\pi$ and the electron charge, respectively. Assuming a circular cross section of the Fermi surface $A_{k}(E_{F})= \pi k_{F}^2$ the corresponding values for the Fermi wave numbers can be calculated for the two crystals, and come out at $k_F$ = 4.4 $\times 10^6$~cm$^{-1}$ and 3.2 $\times 10^6$~cm$^{-1}$, respectively. Next the 2D carrier density $n_{2D}$ can be calculated from the non-spin degenerate relation $n_{2D} = k_{F}^2/4 \pi$ = 1.5 $\times 10^{12}$~cm$^{-2}$ and 0.81 $\times 10^{12}$~cm$^{-2}$, respectively. For a spheroidal Fermi surface with an aspect ratio of 2 these values of $k_F$ would result in a bulk carrier density $n_{3D} = (1/2)\times (2k_F)^3/(3 \pi ^2)$ of $\sim 10^{19}$~cm$^{-3}$ assuming no spin degeneracy. This value exceeds the bulk carrier concentration calculated from the Hall data~\cite{Pan2014} by a factor of 1000.

\begin{figure}
\includegraphics[width=9cm]{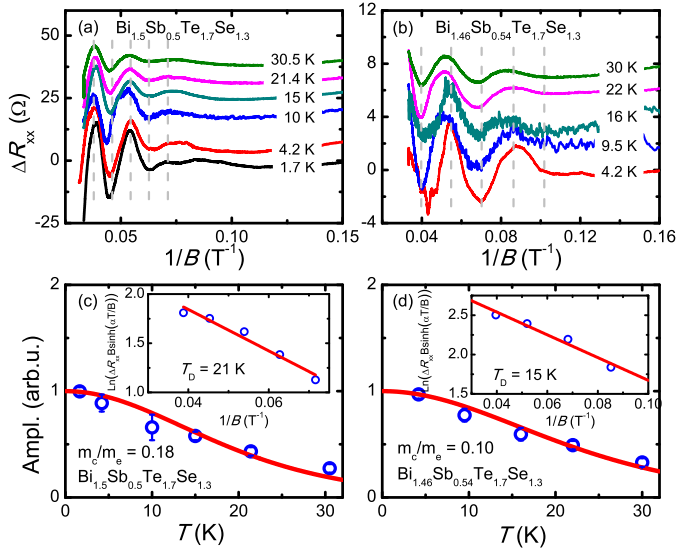}
\caption{Panel (a) and (b): Oscillatory component of the longitudinal resistance $\Delta R_{xx}$ plotted \textit{versus} $1/B$ for BSTS crystals with $(x,y) = (0.50,1.3)$, and $(x,y) = (0.54,1.3)$, respectively, at temperatures in the range 1.7-30~K as indicated, and $\theta = 0^{\circ}$. Curves are offset along the vertical axis for clarity. The vertical dashed lines mark the minima and maxima in $\Delta R_{xx}$. Panel (c) and (d): Thermal damping of the SdH oscillations for $(x,y) = (0.50,1.3)$ and $(x,y) = (0.54,1.3)$, respectively. The cyclotron mass is $m_c = 0.18 m_e$ and $0.10 m_e$, respectively. The insets show the fit to the Dingle damping term at $T = 4.2$~K with the resulting Dingle temperature $T_D = 21$~K and $15$~K, respectively.}
\end{figure}

In order to obtain important information about the transport parameters of the 2D carriers we have measured the temperature variation of the SdH effect for $(x,y) = (0.50,1.3)$ and $(x,y) = (0.54,1.3)$. See Fig.~2(a) and 2(b), respectively. From the thermal damping of the SdH oscillations one can deduce the cyclotron mass, $m_c$, while the amplitude of the SdH oscillations as a function of $B$ allows one to determine the Dingle scattering time, $\tau _D$. The SdH oscillations are analyzed with the standard Lifshitz-Kosevich (LK) expression for 2D charge carriers~\cite{Isihara&Srmcka1986}:
\begin{equation}
\Delta R_{xx} \propto R_{T} R_{D} \cos [2 \pi (\frac{F}{B} - \gamma)],
\end{equation}
where the thermal damping factor $R_T = \frac{\alpha T}{B} / \sinh(\frac{\alpha T}{B})$ with $\alpha = 2 \pi^2 k_B m_c / \hbar e$ and the Dingle damping factor $R_D = \exp(- \alpha T_D / B)$ with the Dingle temperature $T_D = \hbar / 2 \pi k_B \tau_D$. Here $k_B$ is Boltzmann's constant and $\gamma$ is the phase factor. Fits of the measured thermal damping to the LK expression are shown in Fig.~2(c) and 2(d). We extract a cyclotron mass $m_c$ of $0.18 m_e$ and $0.10 m_e$ for $(x,y) = (0.50,1.3)$ and $(x,y) = (0.54,1.3)$, respectively, where $m_e$ is the free electron mass. Combined with the values of $k_F$, derived above, the effective Fermi velocity $v^{\ast} _F \equiv \hbar k_F / m_c $ is calculated and equals 2.8 and 3.6$\times 10^5$~m/s, respectively. The analysis of the Dingle term is shown in the insets to Figs.~2(c) and 2(d) and results in a $T_D$ of 21~K and 15~K and a scattering time $\tau_D$ of 5.8$\times 10^{-14}$~s and 8.4$\times 10^{-14}$~s, for $(x,y) = (0.50,1.3)$ and $(x,y) = (0.54,1.3)$, respectively. By using these values of $v^{\ast}_F$ and $\tau_D$, the mean-free path of the surface carriers, $\ell^{SdH}_{s}$, can be derived from the relation $\ell^{SdH}_{s} = v^{\ast}_F \tau_D$ and amounts to $\sim$16~nm and 30~nm, respectively. Finally, the corresponding surface mobility, $\mu^{SdH}_{s} = e\ell^{SdH}_{s} / \hbar k_F$ is calculated to be $\sim$560 and 1450~cm$^2$/Vs, respectively.

\subsection{Landau level plot and Berry phase}

\begin{figure}
\includegraphics[width=7cm]{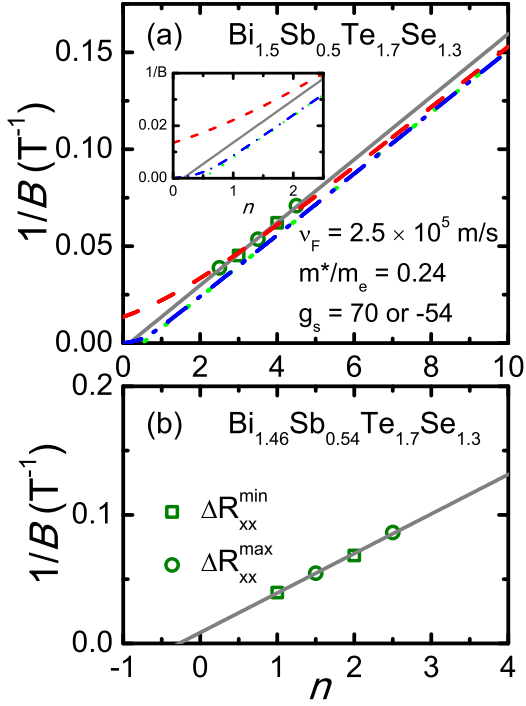}
\caption{Panel (a) and (b): Landau level plot of SdH oscillations in BSTS crystals with $(x,y) = (0.50,1.3)$, and $(x,y) = (0.54,1.3)$, respectively. Minima (green squares) and maxima (green circles) in $\Delta R_{xx}$ correspond to $n$ and $n+1/2$, respectively. In (a) and (b) the grey solid line represents a linear fit with $n_x = 0.16$ and $-0.28$, respectively. In (a) the blue dash-dotted line displays the case of a non-ideal linear dispersion with band parameters $v_F = 2.5\times 10^5$~m/s and effective mass $m^{\ast} = 0.24 m_e$; the red-dashed line includes the Zeeman term with $g_s = 70$ or $-54$; the green dotted-straight line shows the case of the ideal Dirac dispersion with $F = 63$~T. The inset presents a zoom of the LL plot near the origin. For $(x,y) = (0.54,1.3)$ we did not make the full analysis based on eq. 2, since we do not have access to precise values of $m^{\ast}$ and $v_F$. }
\end{figure}

Next, we extract and discuss the Berry phase of the quantum oscillations. The Berry phase $\phi_B = \pi(1-2\gamma)$ can be obtained from the phase factor $\gamma$ in eq.~1 and is $\pi$ for a linear energy dispersion ($\gamma=0$) and zero for a parabolic energy dispersion ($\gamma= 1/2$)~\cite{Mikitik&Sharlai1999}. The standard procedure to extract the phase of the SdH oscillations makes use of a Landau level (LL) plot, in which the LL index $n$ is plotted as a function of $1/B$. In the ideal case $n(1/B)$ is a linear function which extrapolates to $1/B = 0$ at the abscissa $n_x$, where $n_x = 1/2-\gamma$. To construct the LL plot correctly, it is crucial to assign the index $n$ to the correct position in $\Delta R_{xx} (B)$. In our case, for the surface states the Hall resistivity $\rho_{xy} > \rho_{xx}$, which means $\Delta R_{xx}$ has minima (maxima) at integer LL indices $n$ ($n+1/2$)~\cite{Isihara&Srmcka1986}. The corresponding LL plots for $(x,y) = (0.50,1.3)$ and $(x,y) = (0.54,1.3)$ are shown in Figs.~3(a) and 3(b), respectively. A linear least-squares fit (grey solid lines) yields $n_x = 0.16$ and $n_x = -0.28$, and a finite Berry phase $\phi_B = 0.32\pi$ and $-0.56\pi$, respectively. Clearly, these values differ from the value $\pi$ expected for topological surface states.

A Berry phase extracted in this way that deviates from $\pi$ has been reported frequently in other SdH studies on the 3D TI family (Bi,Sb)$_2$(Te,Se)$_3$~\cite{Analytis2010a,Qu2010,Ren2010,Sacépé2011,Xiong2012}. However, obtaining $n_x$ by linear extrapolation is not justified in all cases. While it is appropriate for light-element materials, such as graphene~\cite{Zhang2005}, it is generally not suitable for 3D Bi-based TIs where deviations from the linear dispersion relation, $E(k)$, and the large Zeeman term should be taken into account~\cite{Taskin&Ando2011,Wright&McKenzie2013}. We first investigate the effect of a non-ideal-Dirac $E(k)$. The dispersion relation $E(k)$ for our $(x,y) = (0.50,1.3)$ crystal was directly determined using ARPES. Data measured along the  $\Gamma \rightarrow K$ and $\Gamma \rightarrow M$ high symmetry directions are shown in Fig.~4. In order to fit the energy dispersion of the topological surface state as accurately as possible we need to maximize the number of data points in the occupied part of the energy spectrum. In this sense, the time-dependent energy shift to lower energies observed in BSTS ~\cite{Frantzeskakis2015,Frantzeskakis2015a} and other Bi-based TIs ~\cite{King2011,Zhu2011a} is beneficial. We therefore chose to fit data acquired on a sample which has been maintained for 8~h in a background pressure in the mid $10^{-11}$~mbar range. The data are adequately described by the relation~\cite{Culcer2010,Taskin2011}
\begin{equation}
E(k) = E_{DP} + v_F \hbar k +\frac{\hbar^2}{2 m^{\ast}}k^2,
\end{equation}
where $v_F$ is the Fermi velocity at the Dirac point, $m^{\ast}$ is the effective mass, and $E_{DP}$ is the binding energy of the Dirac point. A least squares fit to eq.~2 gives $v_F = 2.0 \times 10^5$~m/s and $m^{\ast} = 0.24m_e$ for $\Gamma \rightarrow K$ and $v_F = 3.0 \times 10^5$~m/s and $m^{\ast} = 0.25m_e$ for $\Gamma \rightarrow M$. Since the anisotropy is small, we use in the analysis of the LL plot the average values $v_F = 2.5 \times 10^5$~m/s and $m^{\ast} = 0.24m_e$. Following the procedure outlined in Ref.~\cite{Taskin&Ando2011} these band parameters result in the calculated LL plot given by the dashed-dotted blue line in Fig.~3a. In the high-field regime a pronounced curvature towards $n_x =0$ appears, and the LL plot deviates from the ideal linear dispersion relation (straight green dashed line) with $F=63$~T. Clearly, adding the parabolic term in the energy dispersion cannot describe our data properly. Next we include the Zeeman term~\cite{Taskin&Ando2011}, \textit{i.e.} the cyclotron energy $\frac{1}{2} \hbar \omega_c \rightarrow \frac{1}{2} \hbar \omega_c - \frac{1}{2} g_s \mu_B B$. With $g_s = 70$ or $-54$, the LL plot (dashed red line) fits our data well. Such large values of the $g_s$ factor are in-line with those reported for other Bi-based TIs~\cite{Taskin&Ando2011}. Our analysis shows that the linear extrapolation of $n(1/B)$ (solid grey line) does not yield a proper Berry phase. This is due to the large $g_s$ factor and the non-ideal linear $E(k)$. We remark that for higher $n$-values the red and blue LL plots approach each other and coincide for $n > 10$. A linear extrapolation based on this section of the LL plot would yield the true Berry phase. However, to detect quantum oscillations in this regime would require mobilities as large as 2000 cm$^2$/Vs, which these heavily alloyed non-stoichiometric BSTS crystals prepared so far do not offer access to. We conclude that our analysis of the LL plot including the parabolic energy term and Zeeman term is in agreement with topological surface state. Although we did not make a similar full analysis of the LL plot for $(x,y) = (0.54,1.3)$ starting from ARPES data, we note that minor changes in bulk stoichiometry do not affect the ARPES spectra of BSTS~\cite{Frantzeskakis2015}.

\begin{figure}
\includegraphics[height=6cm]{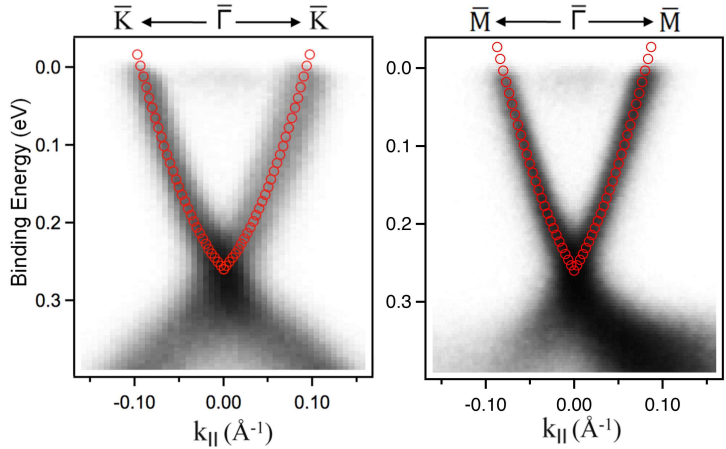}
\caption{Electronic band structure of a Bi$_{1.5}$Sb$_{0.5}$Te$_{1.7}$Se$_{1.3}$ sample acquired by ARPES. The fit using eq.~2 (red open circles) is overlaid on the experimental data (grey-scale). The data are along the $\Gamma \rightarrow K$ (left panel) and $\Gamma \rightarrow M$ (right panel) high-symmetry directions and have been acquired using 27~eV photons and linear horizontal polarization. ARPES has been performed at 16~K.}
\end{figure}

\subsection{Hall resistance}

\begin{figure}
\includegraphics[height=6cm]{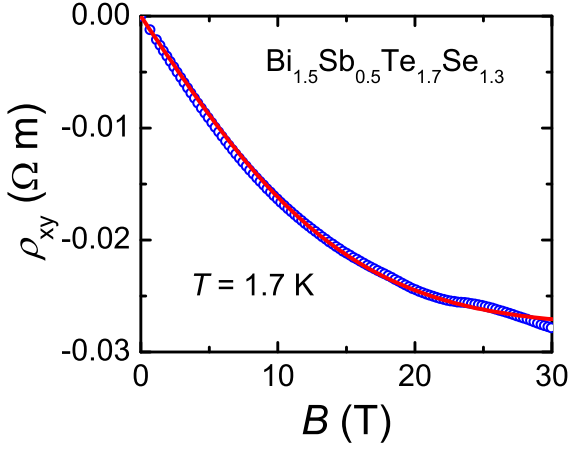}
\caption{Hall resistivity $\rho_{xy}(B)$ of a Bi$_{1.5}$Sb$_{0.5}$Te$_{1.7}$Se$_{1.3}$ crystal of thickness 40~$\mu$m (blue circles) measured at $T = 1.7$~K and $\theta = 0^{\circ}$. The red line shows the fit based on the two-band model described in eq.~3.}
\end{figure}

A full determination of the transport parameters can be made by analyzing the Hall resistivity, $\rho_{xy}$. In Fig.~5 we show $\rho_{xy}$ for the BSTS crystal with $(x,y) = (0.50,1.3)$ and thickness $t = 40 ~\mu$m measured up to 30~T at $T=1.7$~K. Since the surface and bulk-carriers contribute in-parallel to the Hall voltage, we use the standard two-band model~\cite{Xia2013} to fit the data
\begin{equation}
\rho_{xy}(B) = -\frac{B}{e}\frac{(n_b \mu^2_b + n_s \mu^2_s /t)+B^2 \mu^2_b \mu^2_s (n_b + n_s /t)}{(n_b \mu_b + n_s \mu_s /t)^2 + B^2 \mu^2_b \mu^2_s (n_b + n_s /t)^2}
\end{equation}
where $n_b$, $\mu_b$, $n_s$, $\mu_s$ are the bulk carrier density, bulk carrier mobility, surface carrier density and surface carrier mobility, respectively. Here $n_b$ and $\mu _b$ are fit parameters, while $n_s = 1.5 \times 10^{12}$~cm$^{-2}$ and $\mu_s = 560$~cm$^2$ /Vs are taken from the analysis of the SdH oscillations. As shown in Fig.~5, the $\rho_{xy}(B)$ curve (blue circles) is well fitted by the model (red line) and the fit yields $n_b = 2.9 \times 10^{16}$~cm$^{-3}$ and bulk carrier mobility $\mu_b = 15$~cm$^2$/Vs. The surface, $\rho_s$, and bulk resistivity, $\rho_b$, are related via $\rho_s = \rho_{sheet} t = t/ e n_s \mu_s$ with $\rho_{sheet}$ the surface sheet resistivity and $\rho_b = 1/ e n_b \mu_b$, which yields $\rho_s = 29~\Omega$cm and $\rho_b = 14~ \Omega$cm. Therefore, the surface contribution accounts for 32~\% of the total sample conductance according to the formula $\rho^{-1}_s/(\rho^{-1}_s + \rho^{-1}_b)$, which is larger than that reported in Bi$_2$Te$_3$\cite{Qu2010}, Bi$_2$Te$_2$Se\cite{Ren2010} and Bi$_{1.5}$Sb$_{0.5}$Te$_{1.7}$Se$_{1.3}$\cite{Taskin2011}.

\section{Summary}

A magnetotransport study was carried out on low-carrier crystals of the topological insulator Bi$_{2-x}$Sb$_{x}$Te$_{3-y}$Se$_{y}$ with $(x,y) = (0.50,1.3)$ and $(x,y) = (0.54,1.3)$. In high magnetic fields Shubnikov - de Haas oscillations were observed originating from 2D surface states as demonstrated by the angular variation when tilting the sample surface with respect to the field. For $(x,y) = (0.50,1.3)$ the Landau level plot was analyzed with a model incorporating a non-ideal Dirac dispersion that was measured directly using ARPES, and a Zeeman coupling-term with large $g_s$-factor. These effects lead to a shift in the apparent Berry phase extracted from the extrapolated $x$-axis crossing of the linear Landau level plot. Based on the band parameters deduced from ARPES measurements carried out on a sample prepared from the same single-crystalline batch, the SdH oscillations can be attributed to topological surface states with an electron spin $g$-factor $g_s=70$ or $-54$ as fitting parameter in the LL plot model. By combining the carrier density and mobility for the topological surface states from the SdH data with a two-band (bulk + surface) model for the Hall resistivity, the surface contribution to the total electrical transport can be extracted and amounts to around 32~\% in our Bi$_{1.5}$Sb$_{0.5}$Te$_{1.7}$Se$_{1.3}$ crystal with a thickness of 40~$\mu$m.

\vspace{10 mm}
\noindent
\textbf{Acknowledgement}: This work was part of the research program on Topological Insulators funded by FOM (Dutch Foundation for Fundamental Research on Matter).We acknowledge the support of the HFML-RU/FOM, member of the European Magnetic Field Laboratory (EMFL).

\clearpage

\bibliographystyle{elsarticle-num}
\bibliography{RefTI_SdH_SSC_revised}

\end{document}